\documentclass[aps,prl,twocolumn,showpacs,amsmath,floatfix]{revtex4}

\usepackage{graphicx}
\usepackage{dcolumn}
\usepackage{bm}

\begin{document}

\title{Exceptional Points in Atomic Spectra}

\author{Holger Cartarius}
\author{J\"{o}rg Main}
\author{G\"{u}nter Wunner}
\affiliation{Institut f\"{u}r Theoretische Physik 1, Universit\"{a}t Stuttgart,
  70550 Stuttgart, Germany}
\date{\today}

\begin{abstract}
  We report the existence of exceptional points for the hydrogen atom in 
  crossed magnetic and electric fields in numerical calculations. The
  resonances of the system are investigated and it is shown how exceptional
  points can be found by exploiting characteristic properties of the
  degeneracies, which are branch point singularities. A possibility for the
  observation of exceptional points in an experiment with atoms is proposed.
\end{abstract}


\maketitle

The appearance of the coalescence of two eigenstates, so-called exceptional 
points \cite{Kato66}, in physical systems described by non-Hermitian matrices
has attracted growing interest \cite{Hei00,Hei90,Ber98,Dem01,Dem03}.
Typical systems in which
such a phenomenon can occur are open quantum systems with decaying unbound
states. One possibility to describe these open quantum systems are
non-Hermitian Hamiltonians obtained with the complex rotation method 
\cite{Ho83}. In this case, the resonances appear as complex
eigenvalues whose real and imaginary parts are connected with
the energy and the resonance width, respectively.
The eigenvectors have not to be orthogonal in contrast to Hermitian
Hamiltonians describing bound states in quantum mechanics. In
particular, it is possible that the eigenspace for two degenerate eigenvalues
is only one-dimensional, i.e., there is only one linear independent
eigenvector.
If the system of interest depends on a complex parameter $\lambda$ (or two
real parameters), a branch point singularity of two eigenstates can appear at
critical parameter values $\lambda_c$, which are called exceptional points.
Exceptional points have been discovered in a broad variety of physical
systems. Among them are acoustical systems \cite{Shu00}, atoms in optical
lattices \cite{Obe96,Ber98}, and complex atoms in laser fields \cite{Lat95}.
Detailed experiments have been carried out with resonances in
microwave cavities \cite{Phi00,Dem01,Dem03}. However, up to now exceptional
points have not been found in atoms in static external fields. The main
reason is that there is only one parameter in the cases studied most
intensely, viz. atoms either in a magnetic \emph{or} in an electric field.
For the occurrence of exceptional points, the parameter space has to be at
least two-dimensional, i.e., at least two real parameters are required, which
can be represented by crossed magnetic \emph{and} electric fields. Atoms in
static external magnetic and electric fields are fundamental physical systems.
As real quantum systems they are accessible both with experimental and
theoretical methods and have been used for comparisons with semiclassical
theories \cite{Fre02,Rao01,Bar03a}. They are ideally suited to
study the influence of exceptional points on quantum systems. 
E.g., the occurrence of phenomena like Ericson fluctuations in photoionization
spectra has been demonstrated both in numerical studies \cite{Mai92,Mai94} and
experiments \cite{Sta05}. 

In this Letter, we investigate numerically the resonances of the hydrogen atom
in static magnetic and electric fields and report the first detection of
exceptional points in this system. The confirmation of the existence of
exceptional points supplements the richness of phenomena which have been found
in the spectra of atoms in static external fields. Furthermore, we propose
a method which can be used to verify the existence of exceptional points
in experiments. 

\begin{figure}
  \includegraphics[width=\columnwidth]{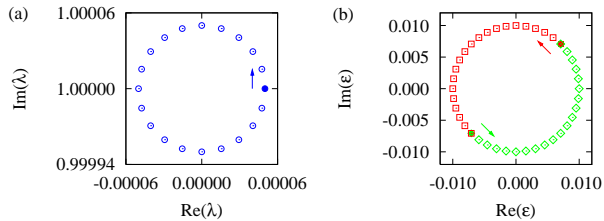}
  \caption{\label{fig:simple_model}(Color online) (a) Circle in parameter
    space $\lambda$ with the exceptional point $\lambda_c = \mathrm{i}$ as
    center point for the simple model \eqref{eq:simple_model}. 
    (b) Eigenvalues $\epsilon_{1,2}$ calculated for the parameter values from
    (a) indicated by squares and diamonds, respectively. In this special
    case, each of the two eigenvalues traverses a semicircle. In (a) and (b),
    the filled symbols represent the first parameter value $\lambda_0$ and
    the corresponding eigenvalues $\epsilon_{1,2}(\lambda_0)$, respectively.
    The arrows point in the direction of progression.}
\end{figure}

Exceptional points can occur in systems which are described by a non-Hermitian
matrix and depend on one complex parameter or two real parameters. At
critical parameter values $\lambda_c$, the eigenvalues and eigenvectors
of the matrix pass through branch point singularities 
\cite{Hei90,Hei91,Ber98}. This can easily be understood by studying the
two-dimensional matrix \cite{Kato66}
\begin{equation}
  \bm{M}(\lambda) = \left ( \begin{array}{rr}
      1 & \lambda \\
      \lambda & -1
    \end{array} \right )
  \label{eq:simple_model}
\end{equation}
with the complex parameter $\lambda$. The eigenvalues 
$\epsilon_1 = \sqrt{1+\lambda^2}$ and $\epsilon_2 = -\sqrt{1+\lambda^2}$
are two branches of one analytic function in $\lambda$. An exceptional point
occurs, e.g., for $\lambda = \mathrm{i}$, where the eigenvectors of the
two degenerate eigenvalues coalesce and the only linearly independent
eigenvector reads $\bm{x}(\lambda = \mathrm{i}) = (1, \mathrm{i})^T$.
The branch point singularity leads to a characteristic behavior of the
corresponding eigenvalues under changes of the parameters. If one chooses
a closed loop in the parameter space and calculates the eigenvalues for a set
of parameter values on this loop, in general, the eigenvalues also traverse a
closed curve. However, when an exceptional point is encircled, two of the
eigenvalues do not traverse a closed loop. The two eigenvalues which degenerate
at the exceptional point are permuted during the traversal of the circle in
parameter space \cite{Kato66}. This can be seen by plotting the paths of the
two eigenvalues in the complex energy plane. After one circle, the first
eigenvalue will travel to the starting point of the second and vice versa,
as illustrated in Fig.~\ref{fig:simple_model}. As a consequence, the path of
one eigenvalue is not closed, if one traversal of the loop in parameter space
is performed. But the path is closed if the parameter space loop is traversed
twice. 

As one would expect, the eigenstates of the system are permuted in the same
way as the corresponding eigenvalues \cite{Kato66}. Additionally, after each
circle around the exceptional point, one of the two eigenvectors changes its
sign \cite{Hei99}, e.g.,
$[\Psi_1, \Psi_2] \overset{\text{circle}}{\to} [\Psi_2, -\Psi_1]$.
This phase change continues with every additional loop. Therefore, four circles
are required to return to the initial states.

In this Letter we investigate exceptional points in the
hydrogen atom in static external fields. The Hamiltonian in atomic units
without relativistic corrections and finite nuclear mass effects \cite{Sch93}
reads
\begin{equation}
  H = \frac{1}{2}\bm{p}^2 - \frac{1}{r} + \frac{1}{2} \gamma L_z
  +\frac{1}{8} \gamma^2 (x^2 + y^2) + f x,
\end{equation}
where $L_z$ is the $z$ component of the angular momentum, and
$\gamma$ and $f$ are the field strengths of the magnetic and electric fields
oriented along the $z$- and $x$-axis, respectively.
The only constants of motion are the
energy and the parity with respect to the $(z=0)$-plane. Here, we
concentrate on states with even $z$-parity.
One possibility to calculate the resonances of the system is the complex
rotation method \cite{Ho83,Del91}. The coordinates of the system $\bm{r}$
are replaced with the complex rotated ones $\bm{r} e^{\mathrm{i} 
  \vartheta}$ in the Hamiltonian and wave functions. The transformation leads
to a complex symmetric matrix representation of the Hamiltonian. Introducing
dilated semiparabolic coordinates (see, e.g., \cite{Mai94}) leads to a
generalized eigenvalue problem of the form
\begin{equation}
  \bm{A}(\gamma,f) \Psi = \kappa \bm{B} \Psi
\end{equation}
with a complex symmetric matrix $\bm{A}(\gamma,f)$, a real symmetric matrix
$\bm{B}$, and the generalized eigenvalue $\kappa = -(1+2 b^4 E)$, where $E$
is the energy and $b$ a complex-valued dilation parameter, which includes the
complex rotation. Above the ionization threshold, resonances are uncovered as
complex energies $E$. The computation of the eigenvalues was performed by
applying the ARPACK library \cite{Leh98} to matrices with dimensions of about
10,000 to 12,300. The algorithm uses the implicitly restarted Arnoldi method
and solves large scale sparse eigenvalue problems efficiently, even for
non-Hermitian matrices. In general, but not at the exceptional points, the
eigenvectors of the resonances can be normalized such that 
$\left < \Psi_i |\bm{B}| \Psi_j \right > = \delta_{ij}$.
The external fields $\gamma$ and $f$ are the two external parameters which
determine the eigenstates. Exceptional points do exist in atomic spectra if
the fields can be chosen in such a way that a coalescence of two states occurs.
The crossed field hydrogen system fulfills all necessary conditions for the
appearance of exceptional points, however, one has to find them in the spectrum
to prove their existence.

In practice, it is very difficult to
look for exceptional points by searching for degeneracies of two complex
eigenvalues. The variation of the parameter values $\gamma$ and $f$ does not
lead to clear indications for degeneracies, and, it is not known in advance
which parameter values are a good choice for starting the search. However, the
method of encircling a point in the  $(\gamma,f)$-plane and searching for
eigenvalues which are permuted has  been very successful. A good choice for
such a closed loop is a ``circle'' in the parameter space of the two field
strengths with a radius $a < 1$ chosen relative to the center $(\gamma_0,f_0)$:
\begin{equation}
  \label{eq:circle_parameter_space}
  \begin{aligned}
    \gamma(\varphi) &= \gamma_0 (1 + a \cos(\varphi)), \\
    f(\varphi) &= f_0 (1 + a \sin(\varphi)).
  \end{aligned}
\end{equation}

The phase change of one of the eigenfunctions after a circle around an
exceptional point can be seen as a sign change of an arbitrary matrix element
$p_{12}=\left < \Psi_1 \right | M \left | \Psi_2 \right >$.
We chose $M = 1$, i.e.,
\begin{equation}
  p_{ij} = \left < \Psi_i \right | \left. \Psi_j \right >,
  \label{eq:matrix_element}
\end{equation}
which can be obtained easily in our calculations and is not diagonal
for the eigenstates which fulfill the orthogonality relation 
$\left < \Psi_i |\bm{B}| \Psi_j \right > = \delta_{ij}$ mentioned above.

\begin{figure}
  \includegraphics[width=0.88\columnwidth]{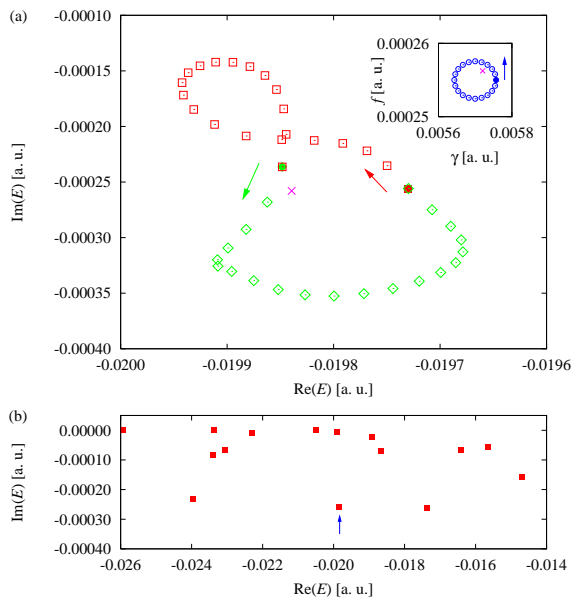}
  \caption{\label{fig:first_ep}(Color online) (a) Paths of the two eigenvalues
    which degenerate at the exceptional point in the complex energy plane. The
    squares and the diamonds represent one of the eigenvalues, respectively.
    Each point of one eigenvalue belongs to a different set 
    of parameters. The path in the field strength parameter space is a circle
    defined in Eq.~\eqref{eq:circle_parameter_space} with $a=0.01$ (see inset).
    The first set of parameters and the corresponding eigenvalues are
    represented by filled symbols. The arrows indicate the direction of
    progression. The cross marks the position of the exceptional point in
    parameter space and the corresponding complex energy of the degenerate
    resonances.
    (b) Resonances in the complex energy plane for the parameter values
    $\gamma = 0.00572$ and $f = 0.000256$ (atomic units) at the exceptional
    point. The position of the two degenerate eigenvalues is marked with an
    arrow.}
\end{figure}

\begin{figure}
  \includegraphics[width=\columnwidth]{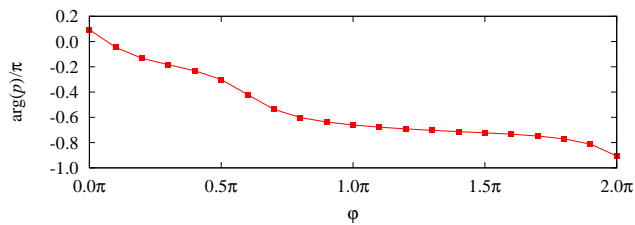}
  \caption{\label{fig:phase} (Color online) Phase of the complex matrix
    element $p$ defined in Eq.~\eqref{eq:matrix_element}. It changes its value
    by $\pi$ during the traversal of the circle in parameter space. The result
    corresponds to the expected sign change.}
\end{figure}

With the method described, exceptional points have been found for the first
time in the spectrum of the hydrogen atom in static external fields.
Fig.~\ref{fig:first_ep}~(a) shows a typical result 
obtained in a numerical calculation. The squares and the diamonds represent one
of the eigenvalues at different field strengths, respectively. In this example,
using 20 steps on the circle in parameter space has been sufficient to obtain
a clear signature of the branch point singularity. The ``radius'' of the
circle according to Eq.~\eqref{eq:circle_parameter_space} was $a = 0.01$.
This value is sufficiently large to have an exceptional point inside the
circular area with a high probability and is small enough to track the paths
of the eigenvalues with a low number of calculations.
Fig.~\ref{fig:first_ep}~(b) shows the position of the degenerate eigenvalues
(marked with an arrow) in the complex energy plane among the resonances in
their vicinity. For the loop shown in Fig.~\ref{fig:first_ep}~(a), the matrix
element $p$ from Eq.~\eqref{eq:matrix_element} was calculated. The phase of
the complex value, plotted in Fig.~\ref{fig:phase}, changes its value by
$\pi$. This clearly indicates a sign change for \emph{one} of the two
eigenstates \cite{Hei99} as mentioned above. 

Examples for parameter values of exceptional points and the position of
the degenerate eigenvalues in the complex energy plane are given in Table
\ref{tab:exceptional_points}. The values were obtained by minimizing the
distance of two eigenvalues which indicated a branch point
singularity. The degeneracy of the two eigenvalues, the existence of only
one eigenvector, and the permutation of eigenvalues were used to identify 
the exceptional points. 

\begin{table}
  \caption{\label{tab:exceptional_points}Examples for exceptional points
    in the spectrum of the hydrogen atom in crossed magnetic ($\gamma$)
    and electric ($f$) fields. All values in atomic units.}
  \begin{ruledtabular}
    \begin{tabular}{D{.}{.}{5}D{.}{.}{6}D{.}{.}{5}D{.}{.}{7}}
      \multicolumn{1}{c}{$\gamma$} & \multicolumn{1}{c}{$f$} 
      & \multicolumn{1}{c}{$\mathrm{Re}(E)$} 
      & \multicolumn{1}{c}{$\mathrm{Im}(E)$} \\
      \hline
      0.00537 & 0.000214 & -0.01884 & -0.0000679 \\
      0.00572 & 0.000256 & -0.01984 & -0.000258 \\
      0.00611 & 0.000256 & -0.01593 & -0.00024 \\
      0.00615 & 0.000265 & -0.0158  & -0.000374 \\
    \end{tabular}
  \end{ruledtabular}
\end{table}

In experiments, the complex eigenvalues cannot be obtained directly. But it is
possible to measure the photoionization cross section and to extract the
energy (real part of a complex eigenvalue) and width (imaginary part) of
the resonances. The cross section can be written in the form \cite{Res75}
\begin{equation}
  \sigma(E) = 4\pi \alpha (E-E_0) \mathrm{Im} \left ( \sum_j \frac{\langle
     \Psi_0^{(\vartheta)} | D | \Psi_j^{(\vartheta)} \rangle^2}{E_j - E} \right ),
  \label{eq:form_cross_section}
\end{equation}
where $\Psi_0^{(\vartheta)}$ is the initial state with energy $E_0$, which is 
supposed to be known. The final state with complex energy $E_j$ is represented 
by $\Psi_j^{(\vartheta)}$, $D$ is the dipole operator for a given polarization,
and $\alpha$ is the fine-structure constant. The superscript $\vartheta$ on
the initial and final states indicates the angle of the complex rotation used
in the calculation, however, it is important to note that the cross section
is independent of that angle in converged spectra.
The Fourier transform $c(t)$ (i.e., the time
signal) of \eqref{eq:form_cross_section} has the form $c(t) = \sum_j a_j
\exp(\mathrm{i}E_j t)$ and, therefore, the complex energies $E_j$
can be obtained using the harmonic inversion method \cite{Bel00} with a high 
precision. In this method, the Fourier transform of the profiles in the
photoionization cross section is used to extract their complex amplitudes
$a_j$ and energies $E_j$ by solving the nonlinear set of equations
\begin{equation}
  c_n = \sum_j a_j \exp(\mathrm{i}E_j t), \quad n = 0,1,\dots,2K-1
  \label{eq:hi_equations}
\end{equation}
for $K$ values $c_n=c(t_n)$ on an equidistant grid. 

\begin{figure}
  \includegraphics[width=0.9\columnwidth]{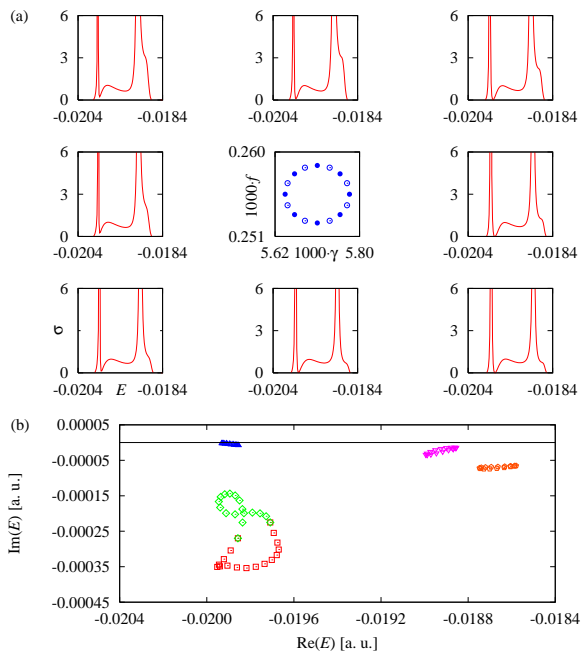}
  \caption{\label{fig:cross_section}(Color online) (a) Photoionization cross
    section of the energy range in which two eigenvalues connected with an
    exceptional point appear. The cross section is shown for eight different
    pairs of parameter values $\gamma$ and $f$ being located on a circle around
    the exceptional point. All values in atomic units.
    (b) Complex energy eigenvalues extracted from the cross sections with the
    harmonic inversion method. Each eigenvalue is drawn with a different
    symbol. Altogether, 16 pairs of parameter values were used. In (a), every
    second cross section is shown. The signature of a branch point singularity
    connected with the two eigenvalues labeled with squares and diamonds,
    respectively, is clearly visible.}
\end{figure}

With this knowledge, a search for exceptional points in an experiment starts
with the measurement of the photoionization cross section for different field
strengths, which are located on a closed loop in the $(\gamma,f)$-space, e.g.,
on a circle of type \eqref{eq:circle_parameter_space}. 
For each measurement, one obtains a spectrum which is used to extract the
complex energies of the resonances. The energy values are drawn in a diagram.
After these steps, the same method which is used to search for branch point
singularities in numerical calculations can be applied to the experimental
results. The diagrams can be used to look for the characteristic open curves
of single eigenvalues. Fig.~\ref{fig:cross_section}~(a) shows an example of a
typical result for the photoionization cross section for different field
strengths which are located on a closed loop in the parameter space. The loop
encircles an exceptional point. Although not experimental but theoretical 
photoionization spectra have been used as input for the calculations, it is
obvious that it is possible to extract the complex energies from
a typical cross section as it is obtained in an experiment. In
Fig.~\ref{fig:cross_section}~(b) one can see the paths of complex eigenvalues
extracted from the spectra. In our example, 16 cross sections were used. This
number is sufficient to give a clear indication of an exceptional point as 
shown in Fig.~\ref{fig:cross_section}~(b).
The eight spectra in Fig.~\ref{fig:cross_section}~(a) look rather similar.
Note that the deviations become more pronounced when the parameter space
radius is increased. This may be helpful for the analysis of experimental
spectra with noise. 

In summary, we have found  the first exceptional points in numerical spectra of
an atom in static external fields. The branch point singularities can be
detected by the permutation of two eigenvalues when an exceptional point is
encircled in parameter space. Further properties of the branch point
singularities can be used to verify their existence. With the harmonic
inversion method, it is possible to extract the complex eigenvalues of
resonances from experimental photoionization cross sections and to detect
exceptional points in experimental data. This opens the way for the 
experimental observation of exceptional points in atomic spectra.

\begin{acknowledgments}
  J.M.\ acknowledges stimulating discussions with W.\ D.\ Heiss.
  This work was supported by DFG. H.C.\ is grateful for the support from the
  LGFG of the Land Baden-W\"urttemberg. 
\end{acknowledgments}


\end{document}